\documentstyle[preprint,aps]{revtex}
\begin{document}
\draft
\title{Quantum measurements with a quantum computer}
\author{C. D'Helon and G.J.Milburn}
\address{Department of Physics, The University of Queensland, QLD 4072, Australia}
\date{\today}
\maketitle
\begin{abstract}
We present a scheme in which an ion trap quantum computer can be 
used to make arbitrarily accurate measurements of the  
quadrature phase variables for the collective vibrational motion of the ion. 
The electronic states of the ion become the `apparatus', and the  
method is based on regarding the `apparatus' as a quantum computer register 
which can be prepared in appropriate states by running a Fourier 
transform algorithm on the data stored within it. The resolution 
of the measurement rises exponentially with the number of ions used.
\end{abstract}
\pacs{ }
Quantum computing offers promise for greatly enhanced efficiency of 
implementation for difficult computational 
problems\cite{Bennett95,DiVincenzo95,Ekert96}. The most
important example to date is the quantum factoring algorithm 
of Shor\cite{Shor94}. Recent advances in quantum error 
correction\cite{Calderbank96,Steane96,Laflamme96}
and fault tolerant computation\cite{Shor96,DiVinShor96} 
indicate that there is nothing in 
principle which will prevent the practical realisation of a quantum 
computer. The most promising suggestion currently is the ion trap realisation
of Cirac and Zoller\cite{CirZol95}. In this scheme, the internal 
electronic states of a string of ions becomes entangled with the lowest 
centre-of-mass vibrational mode of the ion trap. An 
experimental demonstration of a fundamental quantum logic gate in a 
trapped ion realisation was provided by Monroe et al\cite{Monroe96}. 

In this paper we propose that an ion trap quantum computer be viewed as a 
means of making accurate measurements on the centre-of-mass mode of the 
ions. The electronic states of the ions forms an `apparatus' coupled to the 
`system', the vibrational mode. However using the kind of discrete  unitary 
transformations which realise a quantum computing circuit, we can prepare 
the apparatus in a variety of states to facilitate new quantum limited 
measurements on the system. This is in the spirit of the suggestion of 
Wineland and coworkers\cite{Wineland94} that highly entangled states may be used to improve
frequency measurements. 

In  recent experiments on trapped ions\cite{Wineland96} it was possible to 
subject the vibrational motion of the ion to a variety of unitary 
transformations by using sequences of Raman pulses. By selecting the 
Raman detuning carefully, a variety of oscillator states may be excited. 
In some case the unitary transformation of the vibrational motion is 
conditioned on the electronic state of the ion, and may be different for 
the ground and excited states. This conditional dependence was recently 
used to generate a Schr\"{o}dinger cat state for the vibrational mode. In 
other cases the unitary transformation of the oscillator
 is independent of the electronic state.

Consider a single ion laser cooled to the ground state of a
ion trap in the Lamb-Dicke limit. Assume that the ion is illuminated by 
laser fields which are well detuned from resonance. 
Let the interaction Hamiltonian coupling the vibrational and 
 electronic states of a single ion be given by
 \begin{equation}
 \hat{H_I}^{(i)}=\hbar(\alpha_i a^\dagger +\alpha_i^*a)|e\rangle_i\langle e|
 \end{equation}
 where $\alpha_i$ is proportional to a classical laser field acting on the 
 ith-ion in a dispersive regime, $a,a^\dagger$ are the creation and 
 annihilation operators for the vibrational mode of the ion, and 
 $|e\rangle_i$ is the excited electronic state of the ith ion 
 
 If we now generalise to the case of N ions in a linear trap, the 
 electronic state of the system may be described by a 
binary string $S=(S_i=1,0; i=1,N)$ , assuming some ordering of 
the ions, and where a $1,0$ represent the  excited state and ground state 
respectively. Alternatively the state may be described by an integer $k$
for which the binary string is the binary code $k=\sum_{i=1}^NS_i 2^i$. 
There are $2^N$ possible states for the electronic system 
so that $k=0,1,\ldots, K=2^N-1$.  
Assume now that the coupling strengths $\alpha_i$ are adjusted from ion 
to ion so that $\alpha_i=\alpha 2^i$. The  coupling Hamiltonian between the 
 electronic states and the lowest collective vibrational mode of the ions 
 may then be written
\begin{equation}
\hat{H_I}=\hbar(\alpha a^\dagger +\alpha^* a)\hat{\Upsilon}
\end{equation}
where
\begin{equation}
\hat{\Upsilon}=\sum_{k=0}^K k|k\rangle\langle k|
\end{equation}

The phase of $\alpha$ can be adjusted, so for simplicity let us take it to be real. The unitary
operator describing the coupling of the vibrational and  electronic states at the end of a sequence
of pulses is then given by
\begin{equation}
U=e^{ir\hat{x}\hat{\Upsilon}}
\end{equation}
where $\hat{x}=a+a^\dagger$ is the operator describing the in-phase 
component of the collective vibrational amplitude in an interaction 
picture rotating at the ion trap frequency, and $r$ is a real parameter. 

It is clear that the interaction can realise a measurement of 
the vibrational quantity $\hat{x}$. In this interpretation we regard the 
N ion register as an `apparatus' to measure the `system', the vibrational 
mode. However if all the ions are initially 
in the ground state, they will remain in the ground state under this 
unitary interaction as $\hat{\Upsilon}$ commutes with the projection 
operators for each electronic energy eigenstate. In order for the 
interaction to change the state of the ions we must first pre-process the 
electronic register to a state which can be displaced by the unitary 
interaction. The required states $|\bar{k}\rangle$ are eigenstates of an 
operator $\hat{\Phi}$ which is canonically conjugate to $\hat{\Upsilon}$. That 
is to say $\hat{\Upsilon}$ must act as a pure differential operator in 
the $\{|\bar{k}\rangle\ k=0,\ldots,K\}$ basis. 
This basis is simply a discrete Fourier 
transform of the original electronic basis $\{|k\rangle\; k=0,\ldots,K\}$. 
Thus 
\begin{equation}
|\bar{l}\rangle=\frac{1}{\sqrt{K}}\sum_{k=0}^K\exp\left (\frac{2\pi 
ikl}{K+1}\right) |k\rangle
\label{FT}
\end{equation}
It is at this point that we remember the quantum computer interpretation 
of this system. The transformation of Eq (\ref{FT}) may be realised by 
running a Fourier transform algorithm on the electronic register
ions\cite{Beckman96,Dhelon96}. 

Let the initial state of the system be 
\begin{equation}
|\Psi\rangle =|\psi\rangle_v\otimes|0\rangle
\end{equation}
where $|\psi\rangle_v$ is an arbitrary vibrational state and $|0\rangle$ 
indicates all the ions are in the ground state. The {\em first step} is to 
apply a sequence of $\pi/2-$pulses to the ions to place them in a 
symmetric superposition of the ground and excited states. The resulting 
electronic state is precisely the state $|\bar{0}\rangle$, the 0-state in 
the Fourier transform basis. This state is equivalently a uniform 
superposition over all possible electronic energy eigenstates. In the 
{\em second step}, the 
unitary interaction is then implemented to couple the vibrational and 
electronic states. In the {\em third step} an inverse Fourier transform 
is run on the electronic register. At the end of these three steps the 
state of the system is
\begin{equation}
|\Psi^\prime\rangle=\frac{1}{K+1}\sum_{k,l=0}^K\int_{-\infty}^\infty 
dp\phi(p) e^{-\frac{2\pi i kl}{K+1}}|p+rk\rangle_v\otimes|l\rangle
\end{equation}
where $\phi(p)$ is the momentum probability amplitude for the initial 
vibrational state. 

In the final step we readout the state of the electronic register. This 
is very much like the readout of the output register in the Shor 
algorithm. For trapped ions this can be done with very high quantum efficiency 
using quantum jump techniques\cite{Wineland96}. The result of this 
readout is a binary string describing which ions are in the ground state 
and which are in the excited state. Equivalently the result is the integer 
$l$ encoded by this binary string. The (unnormalised) conditional 
vibrational state 
of the system after a readout result, $l$, is 
\begin{equation}
|\tilde{\psi}\rangle_v=\frac{1}{K+1}\sum_{k=0}^K\int_{-\infty}^\infty dp 
\phi(p) e^{-\frac{2\pi i kl}{K+1}}|p+rk\rangle_v
\end{equation}
The probability for this result is found to be 
\begin{equation}
P(l)=\frac{\sqrt{2\pi}}{(K+1)^2}\sum_{k,k^\prime=0}^Ke^{-\frac{2\pi i 
(k-k^\prime)l}{K+1}}\chi(r(k-k^\prime))
\end{equation}
where
\begin{equation}
\chi(k)=\frac{1}{\sqrt{2\pi}}\int_{-\infty}^\infty dx e^{ikx} P(x)
\end{equation}
with $P(x)$ being the initial distribution of the in-phase quadrature 
variable $\hat{x}$, of the vibrational state. Clearly $\chi(k)$ is just 
the characteristic function for the `position' distribution for the 
initial vibrational state. In what follows we will assume that $r=1$.  

The probability distribution for the output 
$l$ is thus seen to a kind of discrete approximation to the probability 
distribution $P(x)$. In the case of a minimum
uncertainty Gaussian state with position variance $\Delta$, and zero mean, the readout distribution
takes the form
\begin{equation}
P(l) =\frac{1}{K+1}\left (1+\sum_{m=1}^K(K+1-m)\cos\left(\frac{2\pi
ml}{K+1}\right )e^{-m^2\Delta/2}\right ).
\end{equation}
 In the limit $\Delta\rightarrow\infty$ we have an effective uniform position 
distribution. In this limit it is easy to see that $P(l)=1$ for all $l$. 
In the opposite limit $\Delta\rightarrow 0$ we have a very well defined 
position at $x=0$, and $P(l)=\delta_{l,0}$. In general the sum can be truncated at
$m_{trunc}=\frac{2\sqrt{2}}{\Delta}$ when subsequent terms contribute less than $e^{-4}$, which we
refer to as the tolerance $\epsilon$. For a large number of ions, the truncation becomes a good
approximation.  In figure \ref{fig1} we illustrate the  intermediate behaviour for
$N= 9$ ions and various values of
$\Delta$,(the tolerance used is $\epsilon=0.01$. In these plots we have reflected the values at $l\
>\ N/2$ to negative values so as to ensure the plot is symmetric around $0$.   We see that as the
number of ions increases we get a better and better  approximation to the true `position'
distribution
$P(x)$.

The dimensionless position $x$ is related to the index $l$ by 
$x=\frac{2\pi}{2^N}l$, 
thus the corresponding position distribution $P^\prime(x)$ obtained from $P(l)$ 
is 
defined over the domain $[0,2\pi)$.
Thus for a fixed number of ions $N$, the range of the uncertainty $\Delta$ in 
the initial 
position distribution $P(x)$ is restricted, if the measured position 
distribution 
$P^\prime(x)$ is to be essentially the same as $P(x)$.
The lower limit imposed on $\Delta$ is determined by the number of ions in the 
register, 
whereas the upper limit on $\Delta$ is independent of the size of the register 
and is 
given by $\Delta_{max}\approx10$.
As the uncertainty approaches this upper limit, the measured electronic 
distribution 
$P(l)$ becomes flat, so that it does not provide any more information about the 
actual 
position distribution $P(x)$.
However, we are assuming that the collective motion of the ions is in the 
Lamb-Dicke 
regime, and hence we require {\em a priori} that the uncertainty in the position is 
bounded by $\Delta x \ll \sqrt{N}/\eta$, where $\eta$ is the Lamb-Dicke 
parameter of 
the centre-of-mass vibrational mode.
Alternatively, if we are given a lower limit for values of $\Delta$, then we can
determine the minimum number of ions $N_{min}$ required to be able to measure 
the 
initial position distribution $P(x)$ accurately, and this is given by 
\begin{equation}
N_{min}\approx 8.14-\frac{1}{2}\log_2 \Delta
\end{equation}
for a tolerance of  $\epsilon =0.01$ in the summation series i.e., this corresponds 
to 
$N_{min}=25$ for $\Delta=10^{-10}$, and $N_{min}=10$ ions for $\Delta=0.1$.

The graphs plotted in Figures \ref{fig2}(a)-(b) demonstrate that the value of 
the uncertainty 
$\Delta x$ in the measured position distribution $P^\prime(x)$ settles to 
the value of the initial uncertainty $\Delta$ as the number of ions increases 
past the appropriate minimum number $N_{min}$, thus indicating that the measured
position distribution $P^\prime(x)$ is approximately equal to $P(x)$.

We have illustrated a way in which an ion trap quantum computer can be 
used to measure the distribution of the in-phase and out-of-phase 
quadrature variables for the collective vibrational motion of the ion. The 
method is based on regarding the ions as a quantum computer register 
which can be prepared in appropriate states by running a Fourier 
transform algorithm on the data stored in the ionic register. 

The first point to note is that we are not restricted to any particular 
quadrature phase variable. As the phase of the laser pulses coupling the 
system and apparatus may be varied we can in principle measure any rotated 
quadrature phase. Given a sufficient set of such distributions a quantum 
tomographic data inversion could be done to reconstruct the initial 
vibrational state.  

The second point to note is that 
as the number of ions increases the states available 
to the computer rises exponentially, providing an increasingly accurate 
readout of a system variable with a continuous spectrum. This generalises 
previous schemes to readout vibrational states using only a single ion; a 
two state apparatus. In that case many repetitions of the measurement 
must be preformed to get an average over particular vibrational 
variables. In our scheme we can do better (albeit at the considerable 
expense of running a quantum algorithm), as we have an  apparatus with 
many more states, and thus one which is better adapted to a vibrational 
degree of freedom.

Finally we note that the scheme is capable of considerable generalisation. 
For example by choosing different couplings between the ion and 
vibrational modes we can get access to many other vibrational variables. 
For example if we choose a sequence of Raman pulses which provided a 
`squeezing' interaction for the vibrational motion, we can measure the 
distribution of eigenstates of the squeeze operator\cite{Chen94}. 
Furthermore, there may 
be other quantum algorithms that realise different measurement schemes. 
\begin{acknowledgments}GJM would like to thank The Institute for Theoretical Physics,
University of California, Santa Barbara, for support during a visit where this work was completed.
\end{acknowledgments}

\begin{figure}
\caption{A plot of the distribution for ion state readouts for an input Gaussian state in the
centre-of-mass coordinate, with two different variances $\Delta$. The number of ions is
9. Dashed: $\Delta=1.0$. Solid:$\Delta=0.1$}
\label{fig1}
\end{figure}
\begin{figure}
\caption{ A plot of the estimated variance in the centre-of-mass coordinate versus ion number for
two different values of the initial coordinate variance. (a)$\Delta=0.1$ (b)$\Delta=1$.  }
\label{fig2}
\end{figure}


\begin{references}
\bibitem[1]{Bennett95}C.H.Bennett, Physics Today,{\bf 48} 24,(1995).
\bibitem[2]{DiVincenzo95}D.P. DiVincenzo, Science, {\bf 269}, 255 (1995).
\bibitem[3]{Ekert96}A.Ekert and R. Jozsa, Rev. Mod. Phys. {\bf 68},733 
(1996).
\bibitem[4]{Shor94}P. Shor, Proc. 35th. Ann. Symp. on Foundations
of Computer Science, IEEE Press, Los Alamitos, 124, (1994).
\bibitem[5]{Calderbank96}A.R.Calderbank and P. Shor, Phys. Rev A {\bf 
54}, 1098, (1996).
\bibitem[6]{Steane96}A.M.Steane, Phys. Rev. Letts. {\bf 77}, 793, (1996).
\bibitem[7]{Laflamme96}R. Laflamme, C. Miquel, J.-P. Paz and W.H. Zurek, 
Phys. Rev. Letts, {\bf 77}, 198, (1996).
\bibitem[8]{Shor96}P.W. Shor in Proc 37th Symp. on Foundations of 
Computer Science, to appear. 
\bibitem[9]{DiVinShor96}D.P. DiVincenzo and P.W. Shor, Phys. Rev. Lett. 
{\bf 77}, 3260, (1996).  
\bibitem[10]{CirZol95}J.I. Cirac and P. Zoller, Phys. Rev. Lett. {\bf 74}, 4714, (1996).
\bibitem[11]{Monroe96}C. Monroe, D. Meekhof, B.E King, W.M. Itano, 
 and D. J. Wineland, Phys. Rev. Lett, {\bf 75}, 4714 (1995). 
\bibitem[12]{Wineland94}D.J.Wineland, J.J. Bollinger, W.M. Itano, D.J. Heinzen, Phys. Rev A {\bf 50} 67, (1994). 
 \bibitem[13]{Wineland96}D.Meekhof, C.Monroe, B.E.King, W.M.Itano and D.J.Wineland,
Phys. Rev. Lett. {\bf 76}, 1796 (1996).
\bibitem[14]{Beckman96}D. Beckman, A. Chari, S. Devabhaktuni and J. 
Preskill, quant-ph/9602016. 
\bibitem[15]{Dhelon96}C. Dhelon and G.J. Milburn, to appear Phys. Rev A 
(1996). 
\bibitem[17]{Chen94}G.J.Milburn, Wenyu Chen and K.R. Jones, Phys. Rev A. 
{\bf 50}, 801 (1994)
\end{references}
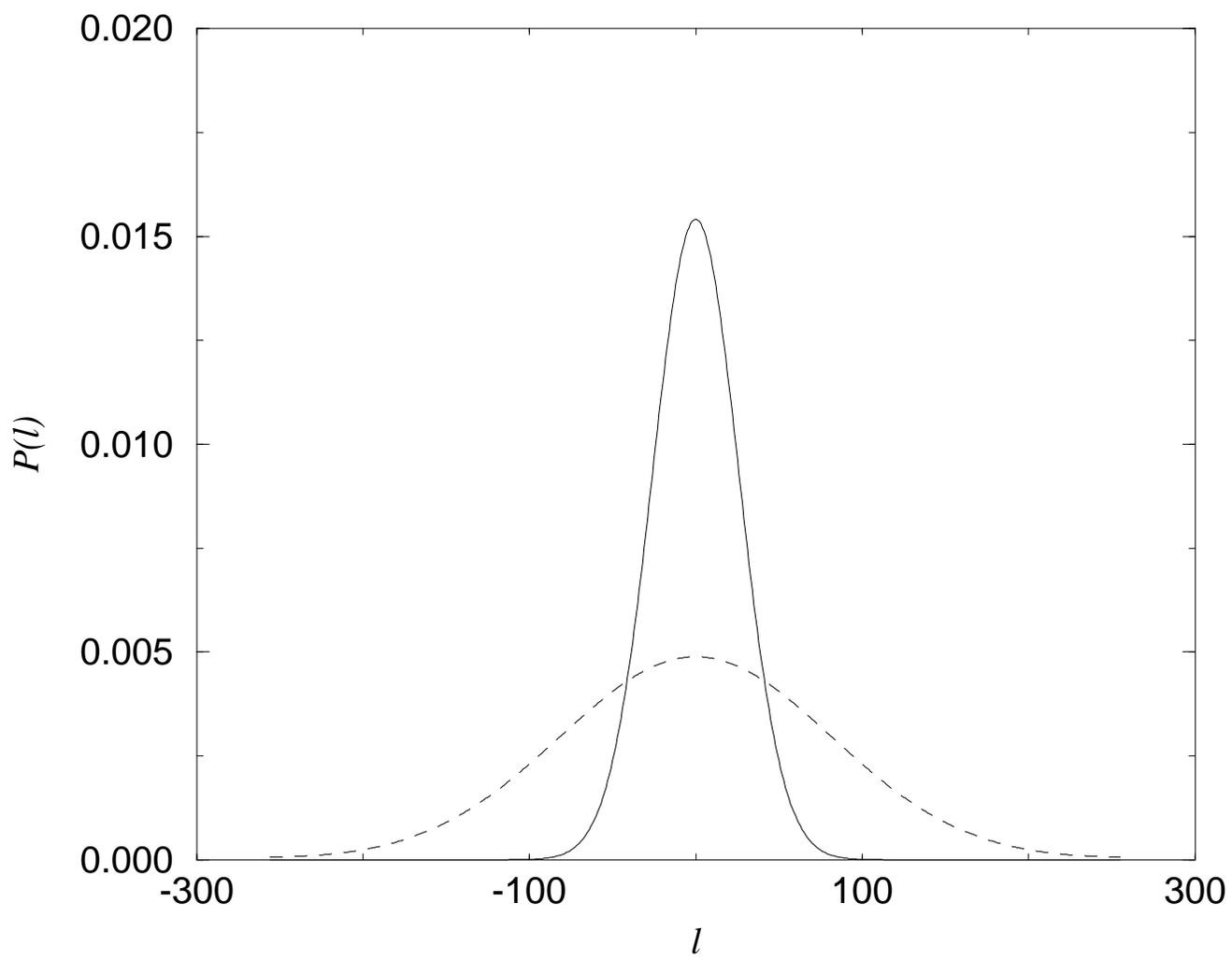
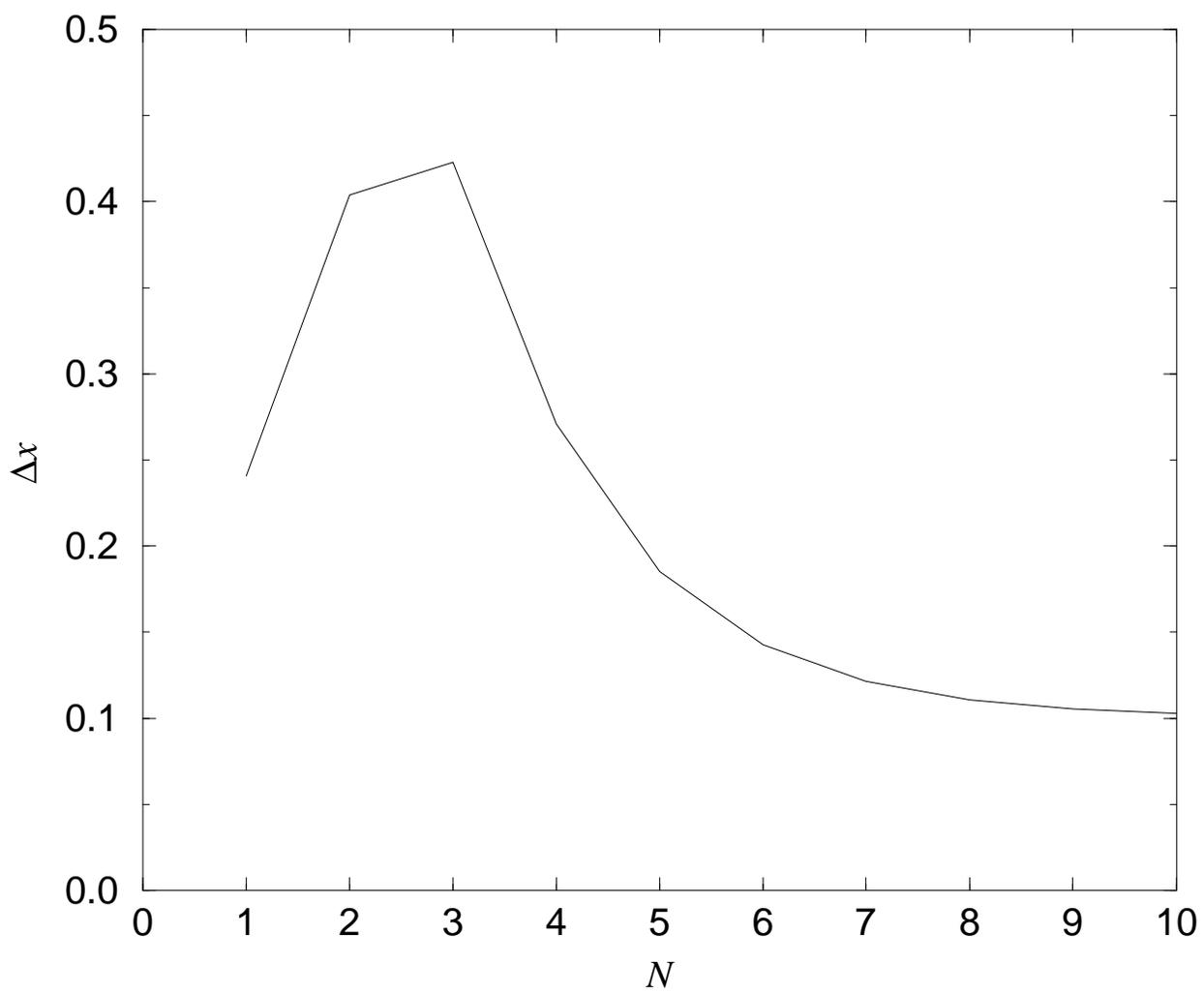
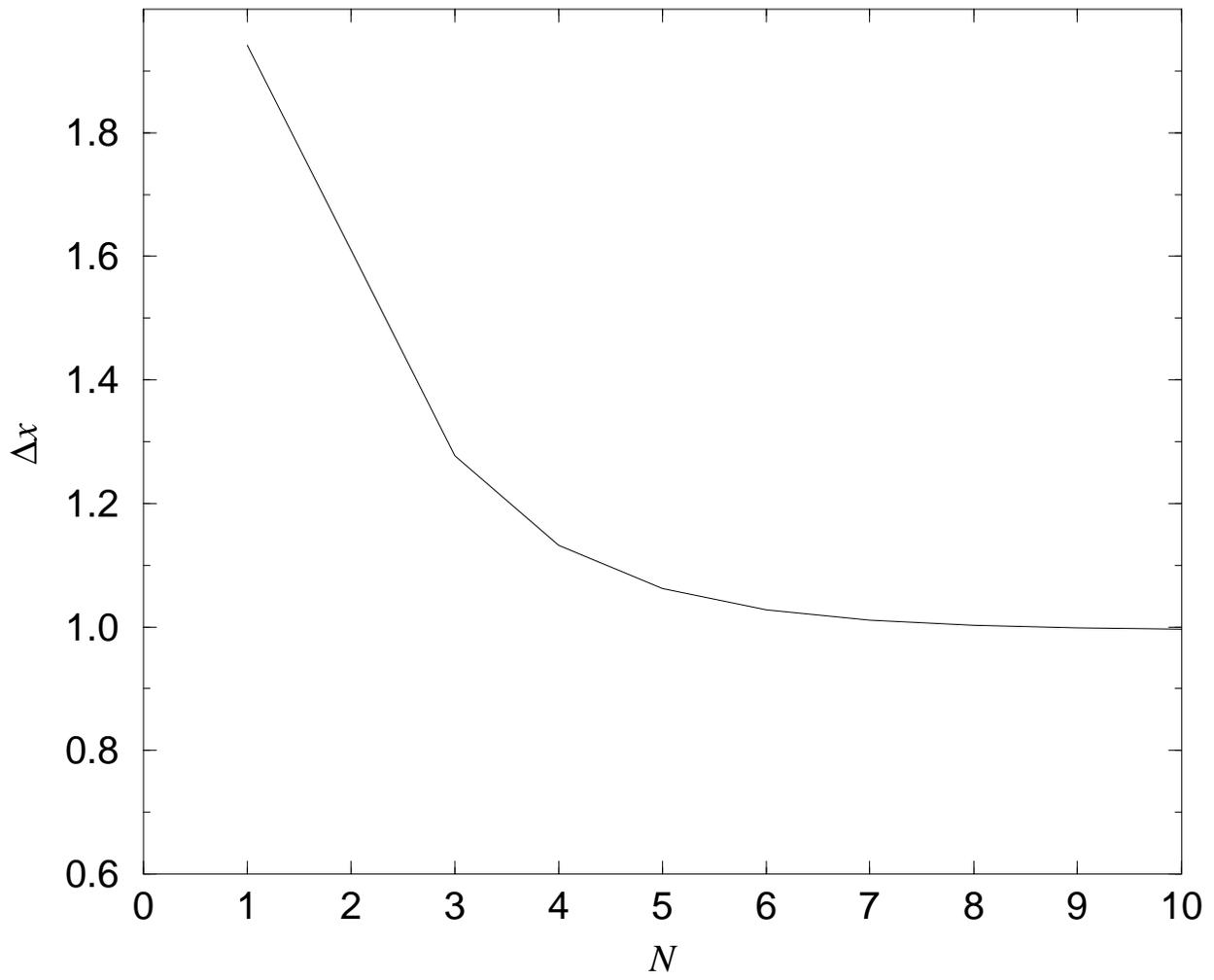
\end{document}